# Sub-cycle time-refraction at optical frequencies

Ohad Segal[†,1], Noa Konforty[†,1], Oded Schiller[1], Maxwell J. Tolchin[2], Jon-Paul Maria[2], Yonatan Plotnik[1] and Mordechai Segev[*,1]

1. Physics Department, Electrical and Computer Engineering department and Solid State Institute, Technion-Israel Institute of Technology, Haifa, Israel.
2. Department of Materials Science Engineering, Pennsylvania State University, USA.
† These authors contributed equally to this work
*msegev@technion.ac.il

**Abstract**:

Large and abrupt variations in the electromagnetic properties of materials lead to dramatic effects: even a single step-like change in the refractive index induces striking phenomena, such as time-refraction and time-reflection. When the refractive index varies periodically in time, multiple time-refractions and -reflections interfere, giving rise to photonic time-crystals (PTCs). Importantly, PTCs display momentum bands separated by gaps in which the modes experience exponential amplification, drawing energy from the modulation in a non-resonant fashion. Ordinary nonlinear optics does not operate in this regime: the material response is either very weak or very slow. One of the immediate consequences is that time-reflection of light at optical frequencies has never been observed in experiments. Here, we experimentally realize an order-unity change in the refractive index occurring at sub-cycle rates, and explore the phenomena emerging from it. By varying the duration of the index change from extending over many cycles to being significantly below a single cycle, we observe that the frequency shift of the time-refraction is enhanced as the index variation occurs faster. Our experiment is the gateway for realizing sharp time-interfaces at optical frequencies, which are the key for experimenting with time-reflection, PTCs and new phenomena expected from light-matter interactions in time-varying media.

The physics of light in materials with time-varying refractive index has been gaining considerable attention in recent years. This is mainly because optical waves can strongly couple to matter, and combining time-varying refractive index and strong light-matter interaction holds great promise for exotic applications such as non-resonant amplification of light, coherent light sources not relying on atomic resonances (*1*), new mechanisms of nonlinear frequency conversion (*2*) and generation of quantum light controlled by the temporal modulation of the index (*3*). These applications and many others (*4–13*) rely on interference between two fundamental processes: time-refraction and time-reflection (*14*, *15*). When an electromagnetic wave is travelling in a homogenous material with time-varying refractive index, its energy (frequency) is altered (as time-translation symmetry is broken by the index variation) but its momentum (wave-vector) is conserved. The conservation of wave-vector implies that, if the refractive index is decreased, the frequency of the wave is blue-shifted (or red-shifted if the index is raised). Time-refraction is characterized by such spectral translation caused by the temporal variation of the refractive index, and was measured in many optical systems (*16–19*). On the other hand, time-reflection amounts to a new wave generated from the index variation, and it is propagating counter to the original wave, with the same frequency as the time-refracted wave. Time-reflection requires refractive index variations that are much harder to produce in experiments because the variations need to occur within a time-frame much shorter than a single cycle of the optical wave and should be of order unity. Because of these extreme conditions, time reflections have not yet been observed in the optical regime, but were thus far observed only with water waves (*20*), cold atoms (*21*) microwaves (*22*, *23*) and in synthetic dimensions (*24*, *25*). Once time-reflections of sufficient amplitudes are observed, one would be able to advance to the next level and explore photonic time crystals (PTCs), which form from periodic variations (in time) of the refractive index. The periodic modulation causes interference between time-refracted and time-reflected waves, that manifests into a band structure with band gaps in momentum (wavenumber) instead of energy (frequency) (*26–28*). The electromagnetic modes of the momentum gaps are exponentially increasing (or decreasing) in their amplitude, drawing energy from the modulation (or transferring energy to it). PTCs were thus far observed at RF frequencies (*27*) and very recently also in microwaves (*29*, *30*). These exponentially increasing modes associated with the momentum bandgap of a PTC were suggested as a source of radiation and non-resonant amplification of light (*1*, *9*), not relying on any atomic resonance and not requiring population inversion. These bandgaps in momentum appear at a region of wavenumbers centered around half the modulation frequency. Modulating a material periodically at twice the period (half the frequency) of an optical wave implies that, for exactly one optical period the refractive index is falling, and for exactly one optical period it is rising; or that either the fall or the rise of the refractive index is sub-cycle. Hence, to achieve interference between time-refractions and time-reflections, to create PTCs and make other exotic applications possible, large (order unity) index changes occurring at sub-cycle time frames are the key.

Until recently, all the experimental observation of order-unity index changes were in the multi-cycle regime (*17*, *18*). Two years ago, the first order-unity index changes occurring within a single optical cycle were reported (*16*, *19*). In these experiments, a probe pulse is launched into a transparent conductive oxide (TCO) sample that is modulated by an intense optical pulse (so-called "modulator") (*17*, *18*, *31–35*). The TCO sample has an epsilon near zero (ENZ) point at the mean wavelength of the probe pulse, and due to the modulator pulse this ENZ point moves, creating a large change (~0.5) in the refractive index for the probe pulse. To directly induce single-cycle index variation, Ref. (*19*) employed a short intense modulator pulse at a shorter wavelength, such that the duration of the modulator pulse is roughly one cycle of the probe period. The TCO samples with their ENZ points near the mean wavelength of the probe were not chosen by coincidence. Rather, these materials display the only mechanism known today that can achieve order-unity variations in the refractive index within a few femtoseconds (fs) while remaining relatively transparent. However, thus far, the single-cycle duration of all experiments was too long to enable significant time-reflections, not to mention the formation of PTCs. For this reason, it is essential to have order-unity index changes occurring with deep sub-cycle time-frames.

Here, we demonstrate exactly that: we observe sub-cycle order-unity variation of the refractive index. We use a sample of doped Cadmium oxide (CdO), which has its ENZ point around $4.3\mu m$ free-space wavelength, and launch a probe pulse with $4\mu m$ mean wavelength (~13fs period) through the sample while it is modulated by an intense 10fs modulator pulse (800nm mean wavelength). From measurements of the transmission and Fresnel-reflection of the probe through the modulated sample, we infer that the refractive index is decreased by ~0.15 during the ~0.77 cycle index variation. We gradually vary the duration of our modulator pulse from 40fs (~3 cycles) to 10fs (sub-cycle) and observe enhanced blue-shift of the time refraction. This experimental result opens the door for realizing sharp time-interfaces, time-reflection, and ultimately photonic time crystals with light-waves at optical frequencies.

As explained above, generating sub-cycle refractive index variations requires several components. The first is an optical probe pulse (incident wave), shown in Fig. 1a. The second is a sample that, while the probe is travelling through it, undergoes an order-unity refractive index variation within a time scale shorter than a single cycle of the probe (few fs), as illustrated in Fig. 1b. The third is that the ENZ point of the TCO sample is at the close vicinity of the (mean) wavelength of the probe pulse. In our experiments, the change in the refractive index in the sample is induced by an intense modulator pulse of duration shorter than a single cycle of the probe (Fig. 1d). Clearly, the second requirement is extremely hard to achieve at optical frequencies. In fact, the only known mechanism that yields sub-cycle index change of order unity is the one associated with TCO materials, when the frequency of the probe pulse is in the vicinity of the ENZ point, where the index change is highly enhanced. In this Letter, we rely on a family of TCO materials that have their ENZ point around $4\mu m$ free-space wavelength.

In our experiments, the probe pulse has a ~100fs duration with energy ~0.01mJ, centered around $4\mu m$ mean wavelength (a single cycle amounts to 13fs), Fig. 1a. This probe pulse is generated by down-converting a pulse from an amplified Ti:Sapphire laser pulse using optical parametric amplifier (OPA) followed by difference frequency generation (DFG). The sample is a $2.7\mu m$ thick doped CdO and the modulator pulse carries energy of ~1mJ centered around $0.8\mu m$ mean wavelength. The modulator pulse is also generated from our amplified Ti:Sapphire laser but it is compressed to <10fs pulses (Fig. 1d) using a hollow core fiber (HCF) filled with Ar gas (*36*). The doping of the CdO sample is chosen such that the ENZ point will be at $4.3\mu m$ (Fig. 1e): just above the probe mean wavelength. This choice of ENZ point allows for order unity refractive index variation in the vicinity of the probe mean wavelength induced by the modulator pulse. Figure 1c illustrates the experimental setup. We measure the Fresnel reflected probe power and the transmitted probe power and spectrum, as a function of the delay between the probe and the modulator pulses, for different pulse durations of the modulator.

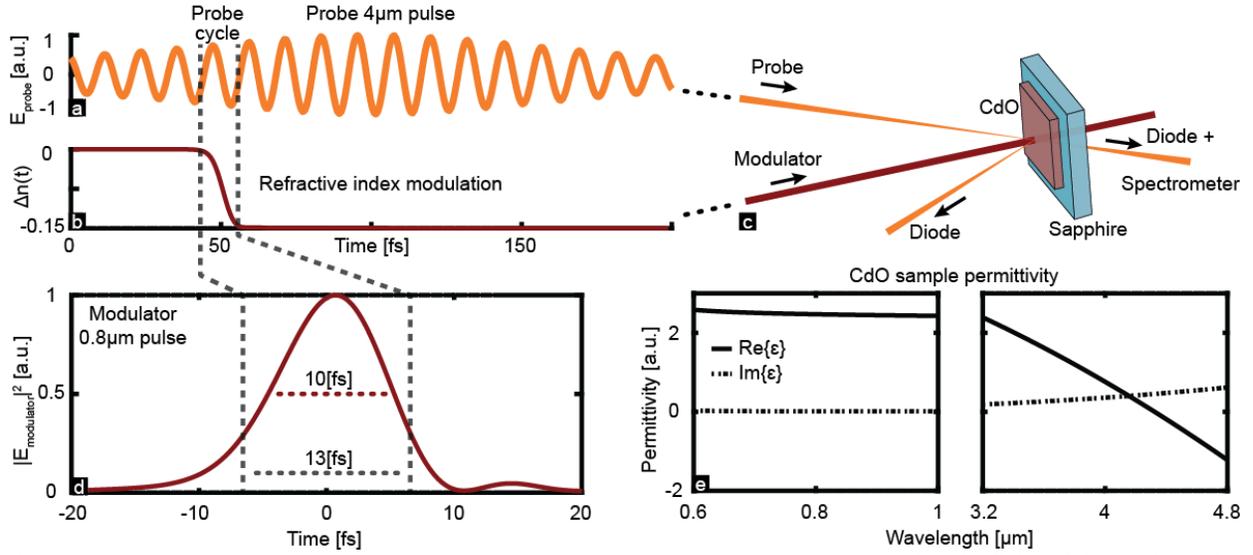

Figure 1: **Experimental scheme of the sub-cycle refractive index variation.** (a) Electric field of the probe pulse ($4\mu m$ mean wavelength), reconstructed using XFROG. (b) Wavefunction of the sub-probe-cycle refractive index variation, calculated by integrating the squared intensity profile of the modulator pulse ($\Delta n(t) \propto -\int_{-\infty}^{t} I_{modulator}^2(\tau)\, d\tau$). (c) Schematic of the experimental modulator-probe setup. The Fresnel reflected probe power and transmitted probe power and spectrum are measured for each modulator-probe delay. (d) Intensity profile of the intense 10fs modulator pulse ($0.8\mu m$ mean wavelength), reconstructed using FROG. (e) Permittivity of the CdO sample in the two relevant wavelength windows, with its ENZ point at $4.3\mu m$.

The permittivity of the CdO sample (Fig. 1e) shows negligible linear absorption at $0.8\mu m$ (modulator mean wavelength). However, the bandgap of CdO is relatively small (~2.4eV (*37–40*)) and therefore, at sufficiently high intensities, the $0.8\mu m$ (~1.5eV modulator) causes significant two photon inter-band transitions. The two-photon inter-band transitions increase the density of conduction electrons, and lead to a shift of the ENZ point to shorter wavelengths. This mechanism decreases the refractive index for the probe pulse.

To unravel the dynamics of the sample permittivity, we scan the delay between modulator pulse and probe pulse, and measure the power of the transmitted and Fresnel-reflected probe (Fig. 2). Figure 2a shows the transmitted probe power as a function of modulator-probe delay, for different durations of the modulator pulse (represented by different colors). To enable comparison between the response at different delays, we set the zero-delay point (for each duration of the modulator pulse) to 10% change of the transmitted probe with respect to transmission measured for the earliest delay. We find that, for all durations of the modulator pulse, the transmission of the probe decreases in the presence of the modulator pulse, in accordance with the shift of the ENZ to shorter wavelengths. Simultaneously, the power of the Fresnel reflected probe (Fig. 2b) increases due to the index change in the sample, implying that the sample becomes more reflective, which also supports the understanding that the ENZ point shift to a shorter wavelength. The relaxation time of both the transmission and the Fresnel reflection back to their original values occur on picoseconds time scales, further verifying that the origin of the observed index change is an inter-band transition, which is always associated with long relaxation times (in contrast to inter-band transitions where the relaxation can be as fast as tens of fs (*19*)).

Next, we examine what happens when we shorten the duration of the modulator pulse while keeping its energy constant. As shown in the rightmost part of Fig. 2b, the Fresnel reflection rises more for shorter modulator pulses, although we make sure to maintain the same energy for the modulator pulses for all temporal widths. From these experiments, we conclude that the observed change in the electromagnetic properties of the sample does not depend on the modulator pulse energy, but depends on the temporal width. Finally, we note that here the change in the refractive index is negative, unlike our previous experiments (*19*) where the index change was positive. From all of these insights, we conclude that the index change here is caused by a multi-photon process (most likely two-photon absorption) caused by the modulator pulse, which increases the density of conduction electrons and leads to a blue-shift of the ENZ point, and consequently to an order-unity negative index change occurring within a sub-cycle time-frame. Using the data presented in Fig. 2, and assuming that the change to the refractive index of the CdO sample can be described by a simple shift of the ENZ wavelength, we estimate that the refractive index variation in our experiments is $\Delta n \approx -0.15$.

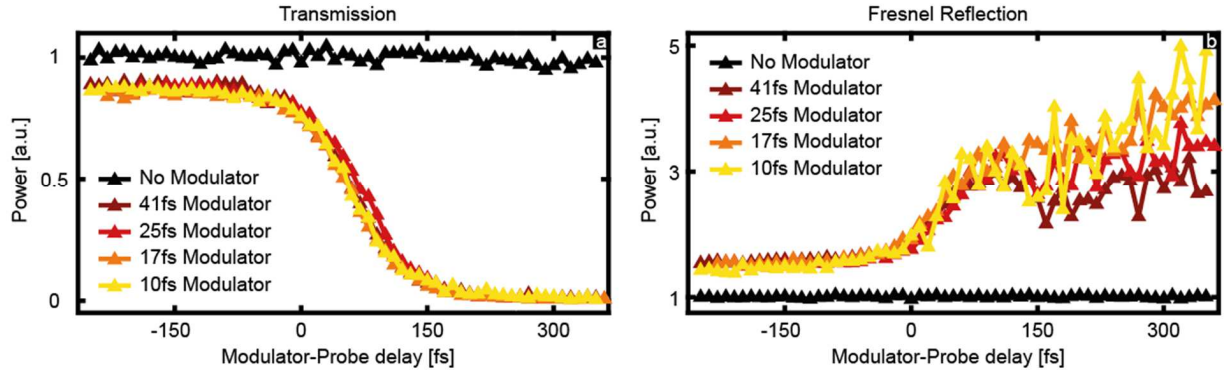

Figure 2: **Transmissivity and reflectivity of the probe beam incident on the CdO sample vs modulator-probe delay and for different durations of the modulator pulse.** **(a)** Transmitted power of the probe vs modulator-probe delay. Negative times correspond to the probe pulse arriving at the sample before the modulator pulse. The modulator pulse causes a decrease in the transmission of the probe pulse. The transmission does not depend on the duration of the modulator pulse. As a reference, black triangles correspond to transmission without any modulator pulse. At negative delays, there is a small difference between the unmodulated and the modulated measurements, due to a weak ghost replica of the probe pulse. **(b)** Fresnel reflection of the probe pulse vs modulator-probe delay (same zero delay as in (a)). The modulator pulse causes an increase in the reflection from the sample. Here, shorter modulator pulses enhance the increase in reflectivity. For comparison, black triangles are without any modulator pulse. At negative delays, there is a small difference between the unmodulated and the modulated measurements due a weak ghost replica of the probe pulse.

A strong signature of the temporal dynamics of the variation in the refractive index is imprinted on the time-refraction – the frequency shift of the transmitted probe pulse. For each duration of the modulator pulse, we measure a spectrogram of the transmitted probe spectrum versus the modulator-probe delay (Fig. 3). As the modulator and probe start to overlap on the sample (between 0fs and 150fs delays in Fig. 3a-d) the spectrum of the probe exhibits a blue shift. This blueshift of the spectrum is another hint of the inter-band transition occurring in the CdO, contrary to intra-band transitions which exhibits mostly a red-shift of the spectrum followed by a blue-shift (*19*). This is because in the current experiments, the modulator pulse increases the density of conduction electrons, which causes a decrease in the index, whereas for intra-band transitions(*16–*

19) the modulator pulse increases the energies of the conduction electrons while keeping their density unchanged, which is manifested in an increase in the refractive index.

Finally, we find that compressing the temporal width of the modulator pulse to be sub-cycle of the probe (Fig. 3d) shows enhanced blueshift, extending to below $3\mu m$ wavelength (13% shorter than without the modulator).

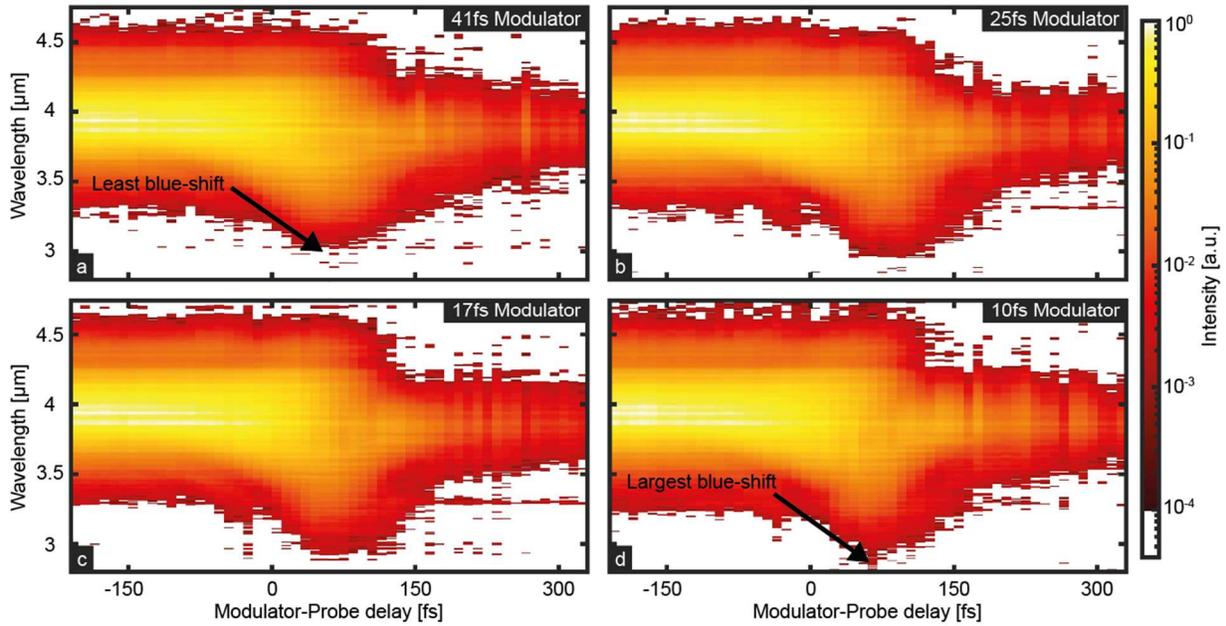

Figure 3: **Spectrograms of the transmitted probe pulse versus modulator-probe delay, for different durations of the modulator pulse. Zero delay is set to be the same as in Fig. 2a. In all panels, new frequencies appear between 0fs and 150fs delays. Namely, the spectrum is blue-shifted to short wavelengths. (a)** 41fs modulator. **(b)** 25fs modulator. **(c)** 17fs modulator. **(d)** 10fs modulator, which is sub-cycle for the probe (where a single cycle is 13fs).

It is instructive to quantify the blueshifts presented in Fig. 3 by plotting the shortest detected wavelength for each modulator-probe delay and for the duration of the modulator pulse (Fig. 4a). This analysis shows that compressing the modulator pulse dramatically enhances the blueshift of the probe pulse. Comparing the shortest measured wavelength over all delays, with and without the index variation, gives a quantitative estimation of spectral shift of the time-refraction for each duration of the modulator pulse (Fig. 4b). The sub-cycle modulator pulse generates the largest frequency shift.

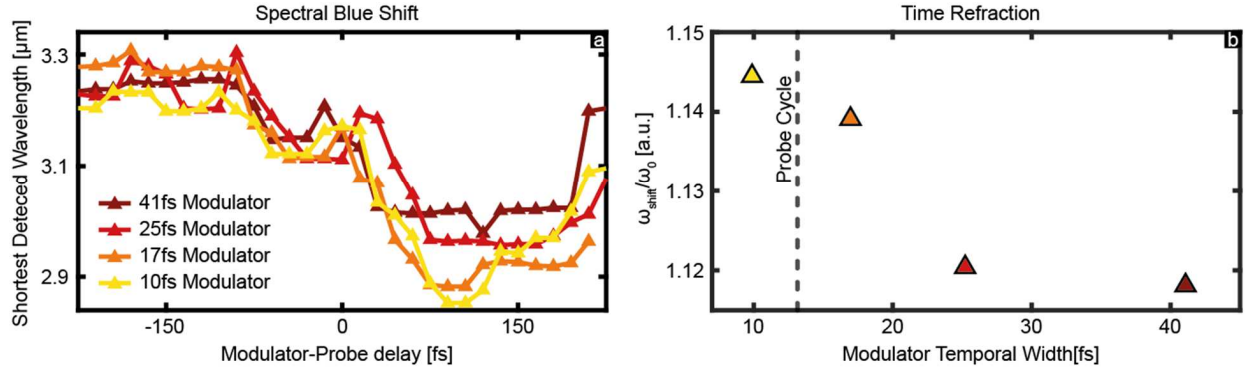

Figure 4: **Spectral blueshift and time-refraction analysis. (a)** Shortest detected wavelength at each delay point of the spectrograms of Fig. 3. Different colors correspond to different durations of the modulator pulse presented in the panels of Fig 3. Shorter modulator pulses create greater blue-shift. **(b)** Estimated spectral shift of the time-refraction calculated from the shortest wavelength of each modulator width in (a) divided by the shortest transmitted wavelength measured without the modulator. The sub-cycle index variation generates the largest spectral shift of the time-refraction.

Our results demonstrate sub-cycle refractive index variation of order unity in an ENZ CdO thin film, driven by intense 10 fs modulator pulses. The induced $\Delta n \approx -0.15$ over <1 optical cycle yields a substantial enhancement of the frequency shifts, validating that the index variation is both large and temporally sharp. This level of ultrafast control is a key prerequisite for observing optical time-reflection, generating sharp time interfaces, and ultimately realizing PTCs at infrared or visible frequencies. The demonstrated platform opens a promising pathway toward experiments with light at optical frequencies in time-varying media. Perhaps most promising future direction is nonlinear frequency conversion in time-varying media, where recent experiments reported observations that cannot be explained by perturbative nonlinear optics theory (*41*, *42*), even though the index changes occurred within several cycles, not yet within an optical sub-cycle. Bringing these effects into the sub-cycle regime will generate a new frontier in nonlinear optics.